\documentclass[preprint,3p,number,sort&compress]{elsarticle}

\makeatletter
\def\ps@pprintTitle{%
 \let\@oddhead\@empty
 \let\@evenhead\@empty
 \def\@oddfoot{\footnotesize\itshape Preprint accepted for publication in 
   Communications in Nonlinear Science and Numerical Simulation\hfill
   March 5, 2017}%
 \let\@evenfoot\@oddfoot}
\makeatother

\usepackage{url}
\usepackage[utf8]{inputenc}
\usepackage[T1]{fontenc}
\usepackage[english]{babel}
\usepackage{amsfonts}
\usepackage{amsmath}
\usepackage{graphicx}
\usepackage{color}
\usepackage[colorlinks=true, allcolors=blue]{hyperref}
\hypersetup{colorlinks = true, allcolors = blue}

\newcommand{\widfig}[2]{\includegraphics[width=#1\columnwidth]{#2}}

\newcommand{\ramuno}{\mathrm{i}\mkern1mu}
\newcommand{\mye}{\mathrm{e}}
\newcommand{\const}{\mathrm{const}}
\newcommand{\mat}[1]{\mbox{\boldmath{$\mathrm{#1}$}}}

\newcommand{\psktot}{q_{\text{tot}}}
\newcommand{\pskprf}{q_{\text{prf}}}
\newcommand{\pskcir}[1]{q_{\text{cir}}^{(#1)}}
\newcommand{\pskrad}[1]{q_{\text{rad}}^{(#1)}}
\newcommand{\regmark}[1]{$\mathbf{#1}$}

\graphicspath{{fig/}}

\begin{document}

\begin{frontmatter}
  \title{Radial and circular synchronization clusters in extended
    starlike network of van der Pol oscillators}

  \author[a]{Pavel V. Kuptsov\corref{cor1}}%
  \cortext[cor1]{Corresponding author}
  \ead{p.kuptsov@rambler.ru}

  \author[a]{Anna V. Kuptsova}%

  \address[a]{Institute of electronics and mechanical engineering, Yuri
    Gagarin State Technical University of Saratov, Politekhnicheskaya
    77, Saratov 410054, Russia}%

  \begin{abstract}
    We consider extended starlike networks where the hub node is
    coupled with several chains of nodes representing star
    rays. Assuming that nodes of the network are occupied by
    nonidentical self-oscillators we study various forms of their
    cluster synchronization. Radial cluster emerges when the nodes are
    synchronized along a ray, while circular cluster is formed by
    nodes without immediate connections but located on identical
    distances to the hub. By its nature the circular synchronization
    is a new manifestation of so called remote synchronization
    [Phys. Rev. E 85 (2012), 026208]. We report its long-range form
    when the synchronized nodes interact through at least three
    intermediate nodes. Forms of long-range remote synchronization are
    elements of scenario of transition to the total synchronization of
    the network. We observe that the far ends of rays synchronize
    first. Then more circular clusters appear involving closer to hub
    nodes. Subsequently the clusters merge and, finally, all network
    become synchronous. Behavior of the extended starlike networks is
    found to be strongly determined by the ray length, while varying
    the number of rays basically affects fine details of a dynamical
    picture. Symmetry of the star also extensively influences the
    dynamics. In an asymmetric star circular cluster mainly vanish in
    favor of radial ones, however, long-range remote synchronization
    survives.
  \end{abstract}

  \begin{keyword}
    networks \sep remote synchronization \sep clusters
\end{keyword}

\end{frontmatter}

\section{Introduction} 

Networks emerging due to some sort of growth process often acquire so
called scale-free structure. It occurs when the growth obeys the
preferential connection rule: already highly connected nodes obtain a
new connection with higher probability compared to those having a
small number of links~\cite{BarabasiAlbert}. This is also known as
Matthew effect or cumulative advantage, i.e., advantage tends to give
rise further advantage and the rich tends to get
richer~\cite{Merton56}. This process results in power law distribution
of node degrees, and the resulting structures are called scale-free
networks.

Scale-free networks, preferential growth and related power laws
attract much of interest of researchers for almost two
decades~\cite{NetwSci2016,Boccaletti2006175,CxNetwTopDynSyn2002}.
Perc~\cite{Matthew2014} surveys Matthew effect in empirical data
ranging from patterns of scientific collaboration and growth features
of socio-technical and biological networks to the evolution of the
most common words and phrases. Scale-free networks formed by
biological cells signaling pathways and regulatory mechanisms are
surveyed by Albert~\cite{Albert4947}. Traffic control in large
networks, self-similarity in traffic behavior and scale-free
characteristic of related complex networks are studied by Dobrescu and
Ionescu~\cite{ScaleFreeTraffic2017}. Barab{\'a}si et
al.~\cite{ScaleFreeNetw} analyze scale-free properties of world-wide
web. Sohn~\cite{IoT2017} introduces scale-free networks as powerful
tools in solving various research problems related to Internet of
Thinks. Bianconi and Rahmede~\cite{HypNetwGeom} consider a
generalization of networks, so called simplicial complexes, and
investigate the nature of the emergent geometry of complex networks in
relation with preferential growth rules and scale-free distributions
of connectivity degree. Bianconi~\cite{BianconiNetwProgress} analyzes
further challenges and perspectives of scale-free network theory.

Populating nodes of a network with oscillators we obtain a dynamical
system with highly nontrivial properties. One of its key effects is
synchronization~\cite{PikRosKurtSyn,Boccaletti2006175, Arenas200893,
  osipov2007synchronization, golubitsky2015recent}. Full, phase or
more subtle forms of synchronization can be observed; it can involve
the whole bunch of nodes or the nodes can form synchronized
clusters~\cite{CxNetwTopDynSyn2002, SyncGraphTopol2005,
  arenas2006synchronization}. Network synchronization phenomena are
extensively studied~\cite{wang2002synchronization, SyncScaleFree2005,
  osipov2007synchronization, Arenas200893}. Jalan et
al.~\cite{ImpLeadClSync2015} analyze cluster synchronization in
presence of a leader. Wang and Chen~\cite{wang2002pinning} investigate
the control of a scale-free dynamical network by applying local
feedback injections to a fraction of network nodes. Covariant Lyapunov
vectors~\cite{CLV2012} and their nonwandering predictable localization
on nodes of scale-free networks of chaotic maps are studied by Kuptsov
and Kuptsova~\cite{NWL2014}.  Li et al.~\cite{QuantSynch17}
investigate the quantum synchronization phenomenon of a scale-free
network constituted by coupled optomechanical systems.

The main motif of scale-free networks is a star. This structure
includes a single hub node connected with several peripheral ones. The
peripheral nodes form star rays. Nodes that belong to different rays
are not connected with each other. Since the stars can be treated as
building blocks for scale-free networks, they attracts a lot of
interest in literature. Chaotic synchronization of oscillator networks
with starlike couplings is considered by Pecora~\cite{Pecora98}. Ma et
al.~\cite{ClustSyncStar2008} study formation of synchronized clusters
in such networks and derive a sufficient condition of existence and
asymptotic stability of a cluster synchronization invariant
manifold. Kuptsov and Kuptsova~\cite{Wild2015} show that starlike
networks of chaotic maps can demonstrate wild multistability, i.e.,
multistability including hardly predictable number of states with
narrow basins of attraction.  For such networks the generalization of
master stability function approach is developed and synchronized
clusters of chaotic nodes are studied~\cite{GMSF}. Chac\'on et
al.~\cite{ImpStarlike16} show that periodic pulses can be used to
effectively control of chaos in starlike networks. Synchronization of
starlike network of fractional order nonlinear oscillators is studied
by Wang and Zhang~\cite{Wang2010}. Hutton and Bose~\cite{SpinStar04}
analyze a quantum system of spins coupled as a starlike network.

One of specific features of starlike networks is so called remote
synchronization. It emerges in networks of nonidentical
self-oscillators as a phase synchronization of peripheral nodes when
the hub is not synchronized with them. Bergner et al.~\cite{RemSyn1}
study this effect for periodic oscillators on compact stars with only
one node per ray. Similar effect in starlike networks of chaotic
oscillators is also known and called relay
synchronization~\cite{RelaySyn06,RelaySyn13,RealySyn16}.  Besides, the
remote synchronization is reported for scale-free
networks~\cite{RemSyn2,RemSyn3}.  Jalan and
Amritkar~\cite{JalanAmritkar2003, JalanAmritkar2005} observes similar
regime for scale-free networks of chaotic maps and call it driven
synchronization.

In this paper we consider extended starlike networks where the hub is
coupled with several chains of nodes of equal
lengths. Figure~\ref{fig:ext_star} demonstrates an extended star with
$R=3$ rays each of $L=2$ nodes. The total number of nodes of an
extended star is $N=RL+1$. Assuming that the nodes are occupied by
nonidentical self-oscillators we study various forms of their cluster
synchronization. The nodes can be synchronized along a ray. We refer
to it as radial synchronization. For example in
Fig.~\ref{fig:ext_star} the radial synchronization occurs when the
node two is synchronized with node three, or it can be nodes four and
five. Also a circular synchronization can be observed when
synchronized clusters are formed by nodes without immediate
connections but located on identical distances to the hub. In
Fig.~\ref{fig:ext_star} the circular synchronization can occur between
near-hub nodes with numbers two, four, and six, and also between
far-end nodes three, five, and seven. By the nature, the circular
synchronization is a new manifestation of the mentioned above remote
synchronization. Unlike the case considered by Bergner et
al.~\cite{RemSyn1}, in this paper a long-range form of the remote
synchronization is reported when the synchronized nodes can
communicate only through at least three node chains. The cluster of
synchronized far ends of rays can appear when other nodes remain
non-synchronized, or when near-hub nodes form another synchronized
cluster. Moreover, far ends can get synchronized with near-hub nodes
too, leaving the hub abandoned. Finally, the total synchronization can
occur when the whole network oscillate synchronously.

\section{The model and synchronization criteria} 

\begin{figure}
  \centering
  \includegraphics[scale=0.75]{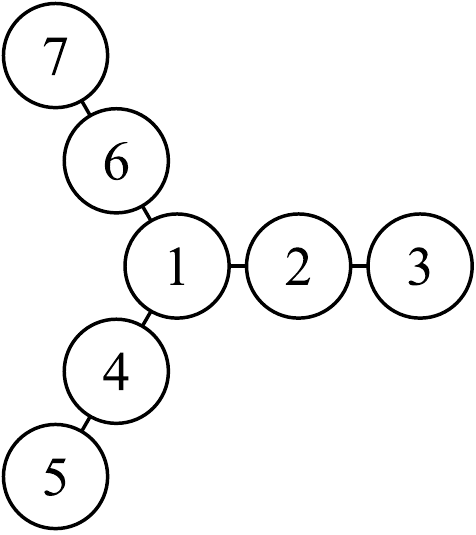}
  \caption{\label{fig:ext_star}Extended starlike network with $R=3$
    rays each of length $L=2$.}
\end{figure}

Consider an extended starlike network of van der Pol oscillators:
\begin{equation}
  \label{eq:netw_vdp}
  \begin{aligned}
    \dot x_n&=y_n,\\
    \dot y_n&=(\mu_n-x_n^2)y_n-\omega_n^2 x_n+
    \epsilon_x s_n+\epsilon_y c_n,
  \end{aligned}
\end{equation}
\begin{equation}
  s_n=\sum_{j=1}^N\frac{a_{nj}}{k_n}x_j-x_n,\;
  c_n=\sum_{j=1}^N\frac{a_{nj}}{k_n}y_j-y_n,\;
  k_n=\sum_{j=1}^Na_{jn},
\end{equation}
where $\omega_n$ and $\mu_n$ are natural frequency and bifurcation
parameter of $n$th oscillator, respectively; $\epsilon_x$ and
$\epsilon_y$ are coupling strengths responsible for reactive and
dissipative coupling, respectively. The coupling structure is
determined by the adjacency matrix $\mat A=\{a_{nj}\}_{n,j=1}^N$. For
the starlike network in Fig.~\ref{fig:ext_star} the matrix $\mat A$
reads:
\begin{equation}
  \label{eq:star_matrix}
  A=
  \begin{pmatrix}
    0 & 1 & 0 & 1 & 0 & 1 & 0\\
    1 & 0 & 1 & 0 & 0 & 0 & 0\\
    0 & 1 & 0 & 0 & 0 & 0 & 0\\
    1 & 0 & 0 & 0 & 1 & 0 & 0\\
    0 & 0 & 0 & 1 & 0 & 0 & 0\\
    1 & 0 & 0 & 0 & 0 & 0 & 1\\
    0 & 0 & 0 & 0 & 0 & 1 & 0
  \end{pmatrix}.
\end{equation}

Bergner et al.~\cite{RemSyn1} detect remote synchronization using
Kuramoto order parameter as an indicator that two nodes oscillate in
phase. For our system~\eqref{eq:netw_vdp} however this characteristic
number is found to depend on parameters in a non-smooth way. To avoid
it we employ the following modified approach. First of all for each
pair of oscillators $m$ and $n$ an array of $j_{\text{max}}$ elements
is initialized whose $j$th cell corresponds to phases from
$j\Delta\phi-\pi$ to $(j+1)\Delta\phi-\pi$, where
$j=0,1,\dots,(j_{\text{max}}-1)$ and $\Delta\phi=2\pi/j_{\text{max}}$
(we use $j_{\text{max}}=100$). When a system moves along a trajectory,
exponentials $\mye^{\ramuno[\phi_m(t)-\phi_n(t)]}$ are computed, where
$\phi_{m,n}(t)=\arctan[y_{m,n}(t)/x_{m,n}(t)]$ are phases of $m$th and
$n$th oscillators, respectively, and accumulated in an array cell with
index $j=\lfloor(\phi_m+\pi)j_{\text{max}}/(2\pi)\rfloor$. After a
long accumulation time content of each cell is averaged and absolute
values are computed. As a result an array of Kuramoto order parameters
separately corresponding to each particular value of $\phi_m$ is
obtained. Finally, average value along the array is computed.  Thus,
the corresponding formula reads
\begin{equation}
  \label{eq:synch_com}
    q_{mn}=
    \left\langle \left|\left\langle 
      \mye^{\ramuno[\phi_m(t)-\phi_n(t)]}
      \right\rangle_t^{\phi_m=\const}\right|\right\rangle_{\phi_m}.
\end{equation}
If oscillations of $m$th and $n$th nodes are totally in-phase,
$q_{mn}=1$. In actual computations we identify the synchronization at
$q_{mn}>0.99$. To distinguish a cluster we will require for all its
node to be pairwise synchronous. For example far-end nodes 3, 5 and 7
of the network in Fig.~\ref{fig:ext_star} are synchronized if
$q_{35}>0.99$, $q_{37}>0.99$, and $q_{57}>0.99$ provided that no one
of them are not synchronized with other nodes.

\section{Charts of synchronization}

\begin{figure}
  \centering
  \widfig{0.32}{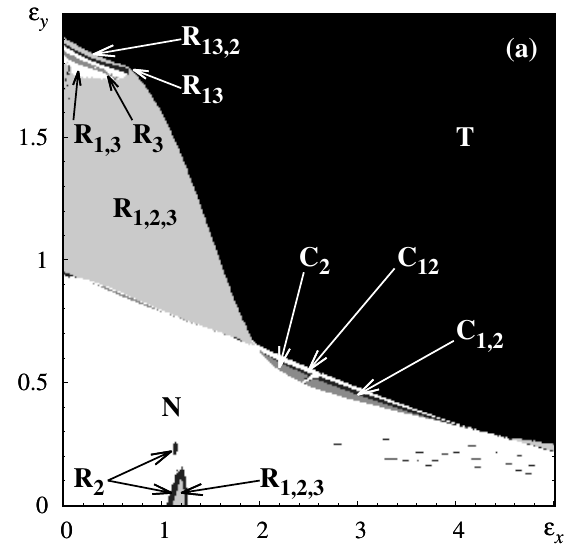}
  \widfig{0.32}{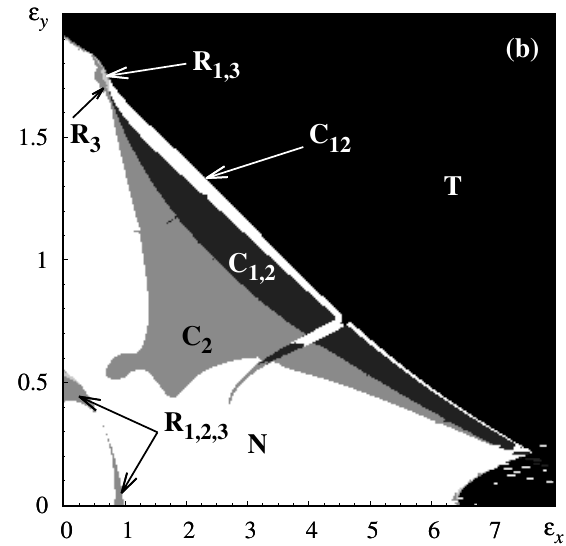}
  \widfig{0.32}{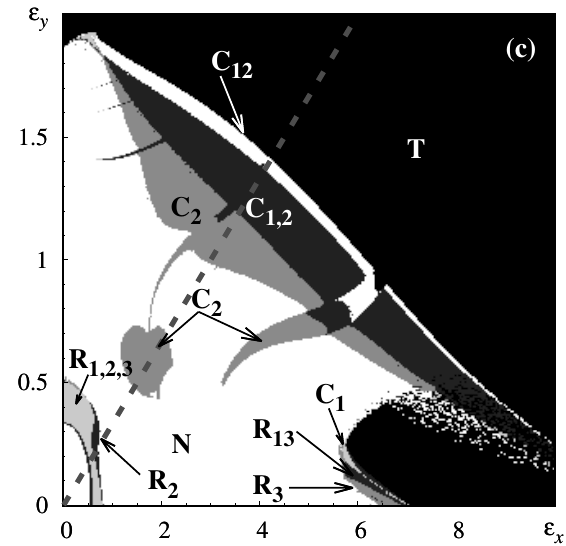}\\
  \widfig{0.32}{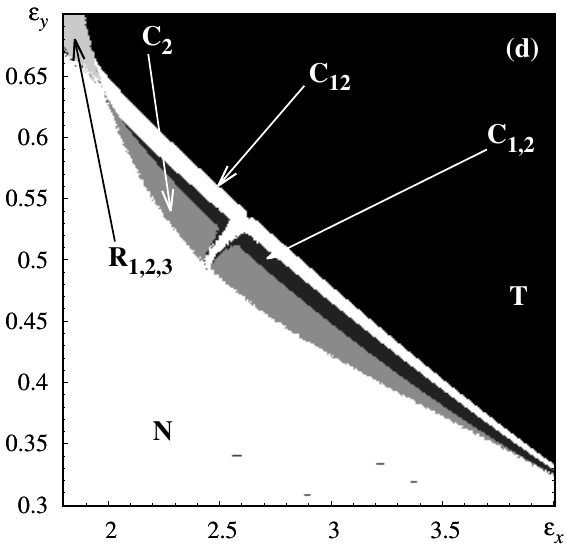}
  \widfig{0.32}{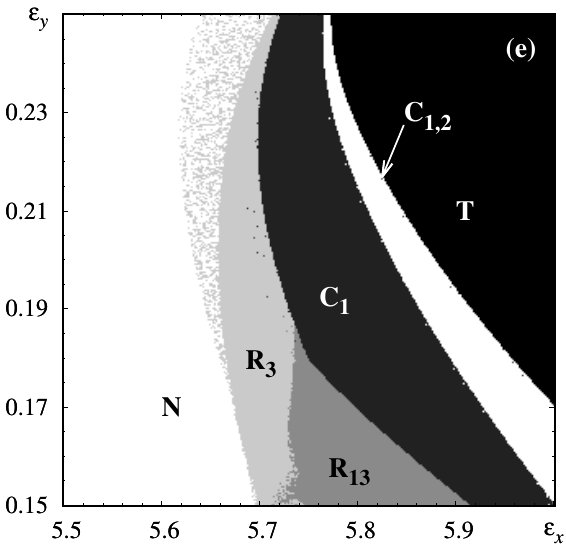}
  \widfig{0.32}{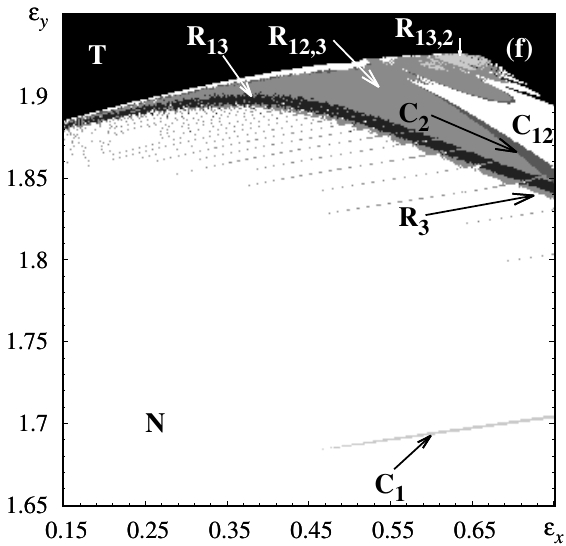}
  \caption{\label{fig:charts}Regimes of extended starlike
    network~\eqref{eq:netw_vdp} whose structure is shown in
    Fig.~\ref{fig:ext_star}. Observe how areas of circular
    (\regmark{C}) and radial (\regmark{R}) clusters separate regimes
    with no synchronization (\regmark{N}) and total synchronization
    (\regmark{T}). (See explanation of subscripts in the text.)
    Natural frequencies of oscillators are given by
    Eq.~\eqref{eq:freq_vals}. Bifurcation parameters in panels (a),
    (b) and (c) can be found in Eqs.~\eqref{eq:charts_a},
    \eqref{eq:charts_b}, and \eqref{eq:charts_c}, respectively. Gray
    dashed line in panel (c) shows where Fig.~\ref{fig:q_lyap} is
    drawn. Panel (d) represents enlarged central area of panel (a). In
    panels (e) and (f) bottom right and top left areas of panel (c)
    are shown.}
\end{figure}

Consider the star with $L=2$, $R=3$, whose structure is shown in
Fig.~\ref{fig:ext_star}. To observe the expected effects of clustering
the natural frequencies of local oscillators have to be 
different. Otherwise the network merely tends to the synchronization
of all nodes (also see Ref.~\cite{RemSyn1} for the corresponding
discussion). We assign the following values:
\begin{equation}
  \label{eq:freq_vals}
  \omega_1=2.5,\;
  \omega_{2,4,6}=\{1.5,\;1.45,\;1.55\},\;
  \omega_{3,5,7}=\{1,\;0.95,\;1.05\},
\end{equation}
i.e., the frequencies strongly decay from the hub to the periphery,
and nodes located on identical distances from the hub have close but
not coinciding frequencies.

Figure~\ref{fig:charts} shows charts of synchronization for the
system~\eqref{eq:netw_vdp},~\eqref{eq:freq_vals}. Letters \regmark{N},
\regmark{C}, \regmark{R}, and \regmark{T} signify areas of no
synchronization, circular synchronization, radial synchronization, and
total synchronization of the whole network, respectively. Subscript
numbers of \regmark{R} denote the rays whose nodes are
synchronized. Rays 1, 2, and 3 include nodes (2, 3), (4, 5), and (6,
7), respectively. Merged subscripts indicate that the corresponding
rays are additionally synchronized with each other, while
non-synchronized rays are separated by commas. Subscripts of
\regmark{C} correspond to the distance from the hub and commas in the
subscripts again separate clusters that are not synchronized with each
other.

In Fig.~\ref{fig:charts}(a) all bifurcation parameter are identical,
\begin{equation}
  \label{eq:charts_a}  
  \mu_{1,\dots,7}=1.
\end{equation}
We see that total synchronization occurs at the top right area of the
chart. Except the total synchronization we also observe a large area
of radial synchronization marked as \regmark{R_{1,2,3}}, i.e., all
nodes along each ray are synchronous but rays are not synchronized
with each other. Also a small island of this regime appears near
$\epsilon_x=1.2$, $\epsilon_y=0$. Regimes \regmark{R_{1,3}} and
\regmark{R_{13}} are observed at the top left part of the figure. The
former indicates synchronization of nodes siting on the first and the
third rays, while the latter represents the case when these rays are
additionally synchronized with each other. Synchronization along one
ray only can also be observed: a very small stripe of \regmark{R_3}
appears at the top left part of the plot, and \regmark{R_2} is
observed in its lower area. Notice that since the star is not fully
symmetric due to non-identical natural frequencies, see
Eq.~\eqref{eq:freq_vals}, the whole set of possible
\regmark{R}-regimes, e.g., \regmark{R_{2,3}} or \regmark{R_{12}}, does
not appear in the plot.

Circular synchronization in Fig.~\ref{fig:charts}(a) is represented by
a narrow stripes \regmark{C_{2}}, \regmark{C_{1,2}}, and
\regmark{C_{12}}. Enlarged this area is shown in
Fig.~\ref{fig:charts}(d). \regmark{C_{2}} indicates synchronization of
far ends only. In area \regmark{C_{1,2}} both far ends and near-hub
nodes are synchronized, but they are not synchronized with each
other. Finally, in area \regmark{C_{12}} all nodes except the hub are
synchronized.

These three regimes are the manifestations of so called remote
synchronization, reported by Bergner et al.~\cite{RemSyn1} for compact
stars, i.e., with $L=1$. According to their explanation it emerges due
to the transfer of signals of synchronized oscillators through the
central node that operates like a filtering coupling line. Unlike
their work, in our case a long-range version of the remote
synchronization can be observed when the nodes interact through three
node chains.

Figure~\ref{fig:charts}(b) illustrates the case when the bifurcation
parameter decays from the hub to far ends:
\begin{equation}
  \label{eq:charts_b}
  \mu_{1}=2,\; \mu_{2,4,6}=1.5,\; \mu_{3,5,7}=1.
\end{equation}
Now areas \regmark{C_{2}}, \regmark{C_{1,2}}, and \regmark{C_{12}} get
larger, while top left \regmark{R}-areas almost vanishes. A tooth-like
area \regmark{R_{1,2,3}} near $\epsilon_x=1.2$, $\epsilon_y=0$
observed in Fig.~\ref{fig:charts}(a) is transformed into a belt around
the origin. This belt is actually surrounded by small borders of
\regmark{R_{1,2}}, \regmark{R_{1,3}}, \regmark{R_{2,3}},
\regmark{R_1}, and \regmark{R_2}, however they are not shown due to
their smallness.

In Fig.~\ref{fig:charts}(c) the gradient of bifurcation parameter is
higher:
\begin{equation}
  \label{eq:charts_c}
  \mu_{1}=3,\; \mu_{2,4,6}=2,\; \mu_{3,5,7}=1.
\end{equation}
We see that areas of the remote synchronization \regmark{C_{2}},
\regmark{C_{1,2}}, and \regmark{C_{12}} are still well
distinguishable, and their boundaries become more complicated: notice
an island and a tongue of \regmark{C_2} in the central area of the
plot; moreover notice a rib-like structure in its top left part.  The
belt \regmark{R_{1,2,3}} also still exists and gets larger.

Unlike panels (a) and (b) of Fig.~\ref{fig:charts} in the panel (c) a
small area \regmark{C_1} appears near $\epsilon_x=6$,
$\epsilon_y=0.2$. This is a one more manifestation of the remote
synchronization where only near-hub nodes are synchronized, while both
the hub and the far ends oscillate independently. In
Fig.~\ref{fig:charts}(e) this area is enlarged.  In addition to areas
\regmark{R_3}, \regmark{R_{13}}, and \regmark{C_1} we can distinguish
here a stripe of \regmark{C_{1,2}}.

A rib-like structure in top left area of Fig.~\ref{fig:charts}(c) is
enlarged in Fig.~\ref{fig:charts}(f). We see that the ribs are formed
by \regmark{R_3} areas, however there is one that corresponds to
\regmark{C_1}. Also notice various \regmark{R}-regimes in the top area
of the plot.

\section{One parameter study of synchronization scenario}

Using $q_{mn}$, see Eq.~\eqref{eq:synch_com} we can estimate degrees
of cluster synchronizations as follows:
\begin{align}
  \label{eq:qrad}  
  \pskrad{r}&=\frac{2}{L(L-1)}\sum_{m=1}^{L-1}\sum_{n=m+1}^L
                        q_{1+(r-1)L+m,1+(r-1)L+n},\;1\leq r\leq R,\\
  \label{eq:qcir}
  \pskcir{\ell}&=\frac{2}{R(R-1)}\sum_{m=1}^{R-1}\sum_{n=m+1}^R
                        q_{L(m-1)+\ell+1,L(n-1)+\ell+1},\;1\leq \ell\leq L,\\
  \label{eq:qprf}  
  \pskprf&=\frac{2}{(N-1)(N-2)}\sum_{m=2}^{N-1}\sum_{n=m+1}^N q_{m,n},\\
  \label{eq:qtot}
  \psktot&=\frac{2}{N(N-1)}\sum_{m=1}^{N-1}\sum_{n=m+1}^N q_{m,n}.
\end{align}
Here $\pskrad{r}$ stands for radial synchronization over $r$th ray,
$\pskcir{\ell}$ shows the circular synchronization of peripherals
nodes located on the distance $\ell$ from the hub, $\pskprf$ shows the
synchronization of all peripheral nodes except the hub, and, finally,
$\psktot$ indicates the synchronization of the whole network. For
example for the star shown in Fig.~\ref{fig:ext_star}
$\pskcir{1}=(q_{2,4}+q_{2,6}+q_{4,6})/3$,
$\pskcir{2}=(q_{3,5}+q_{3,7}+q_{5,7})/3$, $\pskrad{1}=q_{2,3}$,
$\pskrad{2}=q_{4,5}$, $\pskrad{3}=q_{6,7}$.

\begin{figure}
  \centering
  \widfig{0.62}{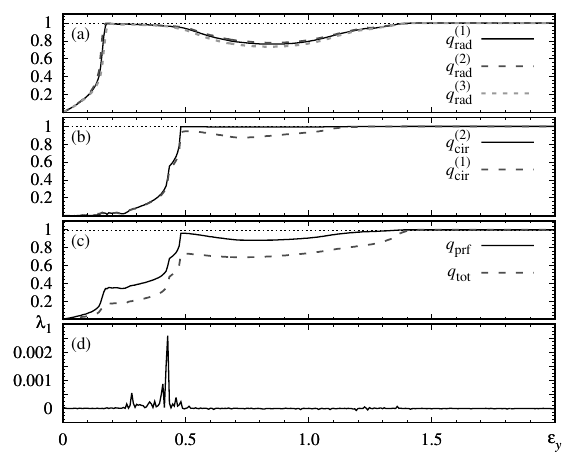}
  \caption{\label{fig:q_lyap}Transition to total synchronization via
    cluster regimes as the coupling strength grows along the gray
    dashed line in Fig.~\ref{fig:charts}(c). Panels (a), (b), and (c)
    shows synchronization criteria~\eqref{eq:qrad}-\eqref{eq:qtot} and
    panel (d) represents the first Lyapunov exponent. Horizontal axis
    corresponds to $\epsilon_y$, $\epsilon_x=3\epsilon_y$, and other
    parameters correspond to Fig.~\ref{fig:charts}(c).}
\end{figure}

Consider a scenario of transition to the total synchronization when
the coupling strengths growth for the case shown in
Fig.~\ref{fig:charts}(c). We will vary $\epsilon_y$ and set
$\epsilon_x=3\epsilon_y$, see the gray dashed line in this figure.
Behavior of the criteria~\eqref{eq:qrad}-\eqref{eq:qtot} along this
line is represented in 
Figs.~\ref{fig:q_lyap}(a-c). Figure~\ref{fig:q_lyap}(d) demonstrates
corresponding values of the first Lyapunov exponent.

Increasing the coupling strength we first observe the emergence of the
radial synchronization at approximately $\epsilon_y=0.18$, see
Fig.~\ref{fig:q_lyap}(a). It corresponds to the belt-like area
\regmark{R_{1,2,3}} in Fig.~\ref{fig:charts}(c). The curves of
$\pskrad{1}$, $\pskrad{2}$ and $\pskrad{3}$ are located very close to
each other but do not coincide entirely because natural frequencies of
node oscillators are slightly different. Oscillations of different
rays are not synchronized with each other. Otherwise $\pskprf$ would
also reach its maximum in Fig.~\ref{fig:q_lyap}(c) which actually does
not take place.

\begin{figure}
  \centering
  \widfig{0.62}{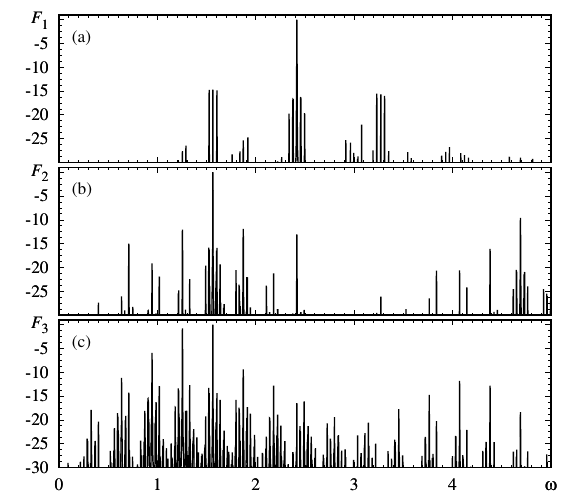}
  \caption{\label{fig:rad1}Quasiperiodic oscillations right below the
    threshold of the radial synchronization indicated by the first
    maximum of $\pskrad{1}$, $\pskrad{2}$, and $\pskrad{3}$ in
    Fig.~\ref{fig:q_lyap}(a). Panels (a), (b) and (c) show Fourier
    spectra of $x_1$, $x_2$ and $x_3$, respectively.
    $\epsilon_y=0.16$. Quite different forms of spectra indicate an
    absence of synchronization.}
\end{figure}

Figure~\ref{fig:rad1} shows Fourier spectra right below the point
where $\pskrad{r}$ attains the maxima in Fig.~\ref{fig:q_lyap}(a). The
first ray, i.e., nodes 1, 2, and 3, is represented. The spectra
consist of series of isolated spikes that corresponds to a rich
quasiperiodicity. The quasiperiodicity is also confirmed in
Fig.~\ref{fig:q_lyap}(c) where the respective first Lyapunov exponent
is zero. Quite different forms of spectra for $x_1$, $x_2$, and $x_3$,
see Figs.~\ref{fig:rad1}(a), (b), and (c), respectively, is an obvious
consequence of oscillators parameters mismatch.

\begin{figure}
  \centering
  \widfig{0.62}{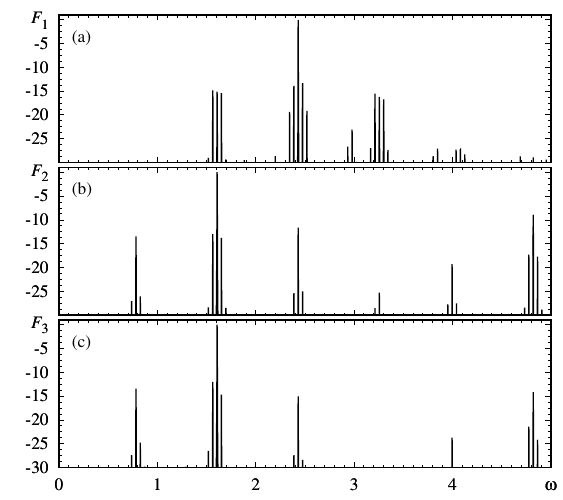}
  \caption{\label{fig:rad2}Quasiperiodicity at the point of emergence
    of the radial synchronization at $\epsilon_y=0.18$, see the first
    maximum of $\pskrad{1}$, $\pskrad{2}$, and $\pskrad{3}$ in
    Fig.~\ref{fig:q_lyap}(a). Panels (a), (b) and (c) represent
    Fourier spectra of $x_1$, $x_2$ and $x_3$, respectively.  Due to
    the synchronization the spectra for nodes 2 and 3 share almost all
    harmonics. Non-synchronized hub, i.e, node 1, though reproduces
    some harmonics, has quite different dominating frequency.}
\end{figure}

The maximum of $\pskrad{1}$ at $\epsilon_y\approx 0.18$ in
Fig.~\ref{fig:q_lyap}(a) indicates synchronization of the nodes 2 and
3. Figure~\ref{fig:rad2}(b) and (c) shows the corresponding Fourier
spectra of these nodes. Observe their very high similarity. The
spectra, however do not fully coincide since phase synchronous
oscillations can be different in their detail structure. The forms of
the spectra again, as in Fig.~\ref{fig:rad1}, indicate the
quasiperiodic nature of the oscillations, and zero of the first
Lyapunov exponent in Fig.~\ref{fig:q_lyap}(d) indicates the
same. Comparing the spectra of peripheral nodes in
Figs~\ref{fig:rad2}(b) and (c) with the hub spectrum represented in
Fig~\ref{fig:rad2}(a), we see that all of them share a large number of
harmonics. But the dominating frequency of the hub is different so
that it remains independent from the peripheral nodes.

\begin{figure}
  \centering
  \widfig{0.62}{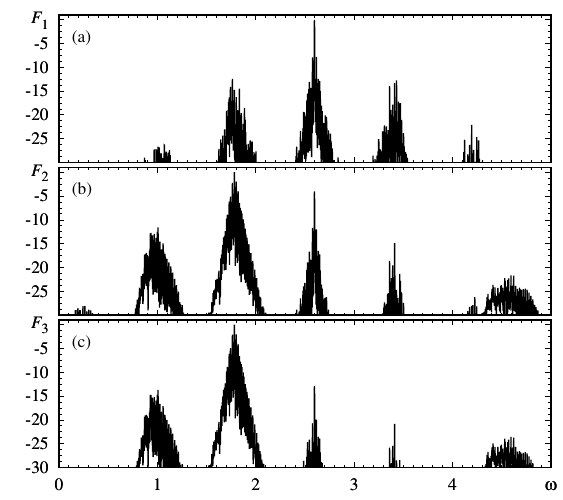}
  \caption{\label{fig:rad3}Chaos emerged as result of destruction of
    radial synchronization. Coupling parameter $\epsilon_y=0.42$
    corresponds to the highest spike of $\lambda_1$ in
    Fig.~\ref{fig:q_lyap}(d).  Panels (a), (b) and (c) represent
    Fourier spectra of $x_1$, $x_2$ and $x_3$, respectively. The
    spectra for nodes 2 and 3 are still similar, i.e., despite the
    destruction of the radial synchronization, some degree of
    coherence survives.}
\end{figure}

Further growth of $\epsilon_y$ results in decay of $\pskrad{r}$, see
Fig.~\ref{fig:q_lyap}(a), indicating the destruction of the radial
synchronization. Corresponding Lyapunov exponent in
Fig.~\ref{fig:q_lyap}(d) becomes positive. This is a footprint of the
transition to chaos via destruction of multidimensional torus. This
scenario is called as Landau-Hopf transition to
chaos~\cite{Tori3D1988,Tori3D1990,Tori3D1993,Tori3D2007,Tori3D2009,Tori3D2011,Tori3D2013,
  Tori4D1985,Tori4D2008,Tori5D2013}. Figure~\ref{fig:rad3} shows
Fourier spectra when the first Lyapunov has the maximum at
$\epsilon_y\approx 0.42$. Instead of the separated spikes observed in
Fig.~\ref{fig:rad2}, the spectra are now continuous that is a specific
feature of chaos. The spectra of $x_2$ and $x_3$, see
Figs.~\ref{fig:rad2}(b) and (c), respectively, are still similar to
each other. The corresponding value of $\pskrad{1}$ is accordingly
close to one. It means that despite the destruction of the radial
synchronization, some degree of coherence of oscillations survives.

\begin{figure}
  \centering
  \widfig{0.62}{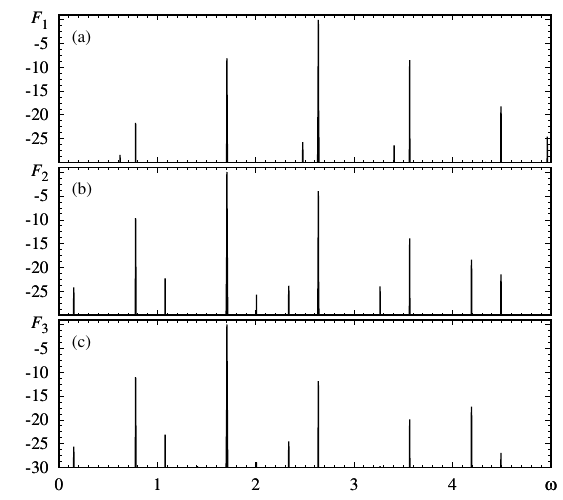}
  \caption{\label{fig:rad4}Reborn quasiperiodicity beyond chaotic
    regime and transition to circular synchronization of far ends at
    $\epsilon_y=0.5$, see Fig.~\ref{fig:q_lyap}(b) where
    $\pskcir{2}=1$ at this point. Panels (a), (b) and (c) represent
    Fourier spectra of $x_1$, $x_2$ and $x_3$,
    respectively. Synchronization of far end is provided by
    information transfer through central nodes: observe that the
    essential harmonics of the far-end node 3 appear also in spectra
    of nodes 1 and 2.}
\end{figure}

Subsequent growth of $\epsilon_y$ results in the destruction of chaos
and reappearance of the quasiperiodicity, see Fourier spectra in
Fig.~\ref{fig:rad4} plotted at $\epsilon_y=0.5$. In comparison with
the situation below the chaotic window, see Fig.~\ref{fig:rad2}, the
reborn quasiperiodicity has much less number of independent
frequencies. The reborn quasiperiodicity appears due to the transition
to circular synchronization of far-end nodes 3, 5, and
7. Figure~\ref{fig:rad4} corresponds to the edge of the island
\regmark{C_2} in Fig.~\ref{fig:charts}(c). This is a manifestation of
long-range remote synchronization. The short-range version of this
effect for compact stars with $L=1$ is reported by Bergner et
al.~\cite{RemSyn1}. This effect occurs since intermediate nodes
operate like filtering communication lines~\cite{RemSyn1}. In our case
both the hub 1 and the near-hub nodes 2, 4, 6 operate like this. This
is illustrated in Fig.~\ref{fig:rad4} that shows how the essential
harmonics of the far-end node 3 appear also in spectra of oscillations
in the central nodes 1 and 2.

\begin{figure}
  \centering
  \widfig{0.62}{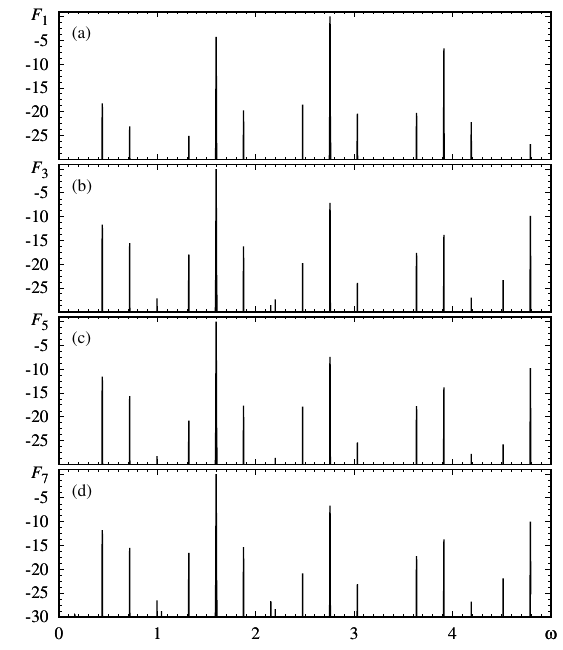}
  \caption{\label{fig:cir1}Circular synchronizations of far ends at
    $\epsilon_y=1$. Panels (a),(b),(c), and (d) represent Fourier
    spectra of nodes $x_1$, $x_3$, $x_5$ and $x_7$,
    respectively. Observe almost identical spectra of far end nodes 3,
    5, and 7. The essential harmonics of these nodes also present in
    the hub spectrum, but the dominating frequency differs.}
\end{figure}

Fourier spectra for the edge of the main area \regmark{C_2} is shown
in Fig.~\ref{fig:cir1}. Observe almost identical spectra in the panels
(b), (c) and (d) corresponding to the synchronized nodes 3, 5 and
7. Notice that the essential harmonics of these nodes are
present in the hub oscillations, panel (a), but the hub dominating
frequency differs from that of far ends.

\begin{figure}
  \centering
  \widfig{0.62}{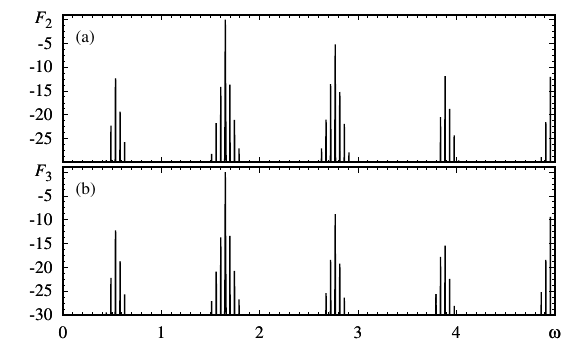}
  \caption{\label{fig:cir2}Independent clusters of near-hub and
    far-end nodes at $\epsilon_y=1.2$. Panels (a) and (b) show Fourier
    spectra of $x_2$, $x_3$, respectively. Though nodes 2 and 3 belong
    to clusters that are not synchronized with each other, their
    spectra are highly similar. This is explained by a twofold role of
    the near-hub node 2: in addition to its own oscillations, it
    provides the coupling line for the far-end node 3.}
\end{figure}

Further increment of $\epsilon_y$ results in the emergence of one more
form of the remote synchronization. In addition to far-end cluster
near-hub nodes are also get synchronized. In Fig.~\ref{fig:q_lyap}(b)
$\pskcir{1}=1$ at $\epsilon_y\approx 1.1$. Since $\pskprf<1$ in
Fig.~\ref{fig:q_lyap}(c), far-end and near-hub clusters oscillate
independently. Figure~\ref{fig:cir2} compares Fourier spectra of the
nodes 2 and 3 at $\epsilon_y=1.2$ when they are involved into
different synchronization clusters, see area \regmark{C_{1,2}} in
Fig.~\ref{fig:charts}(c). Though the dominating frequencies are the
same, the amplitudes of other harmonics are different. The high
similarity of spectra contents is explained by a twofold role of the
near-hub nodes in this case. In addition to their own oscillations
they provide the coupling line for far ends.

Subsequent growth of $\epsilon_y$ results in the synchronization of
both circular clusters, so that all nodes except the hub are
synchronized. In Fig.~\ref{fig:q_lyap}(c) $\pskprf=1$ at
$\epsilon_y\approx 1.3$. Obviously this is the third observed form of
the remote synchronization. The hub gets involved into synchronization
at $\epsilon_y\approx 1.4$, see Fig.~\ref{fig:q_lyap}(c), so that the
whole network becomes synchronized.

\section{\label{sec:larger_stars}Charts of sychronization for lager
  stars}

\begin{figure}
  \centering
  a)~\includegraphics[scale=0.75]{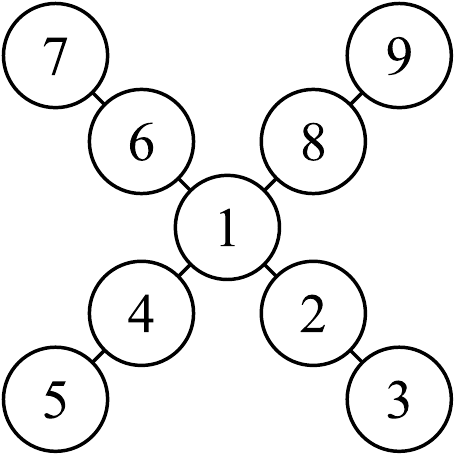}
  \hfill
  b)~\includegraphics[scale=0.75]{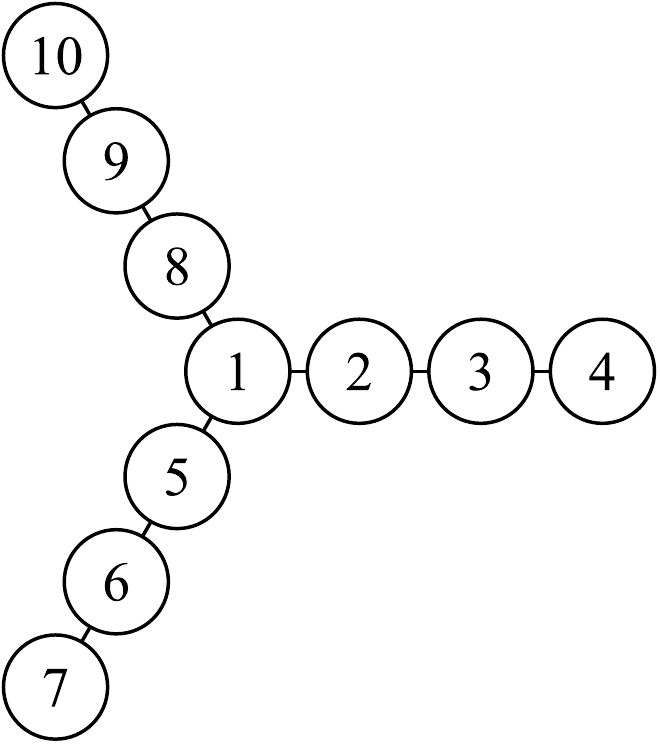}
  \hfill  
  c)~\includegraphics[scale=0.75]{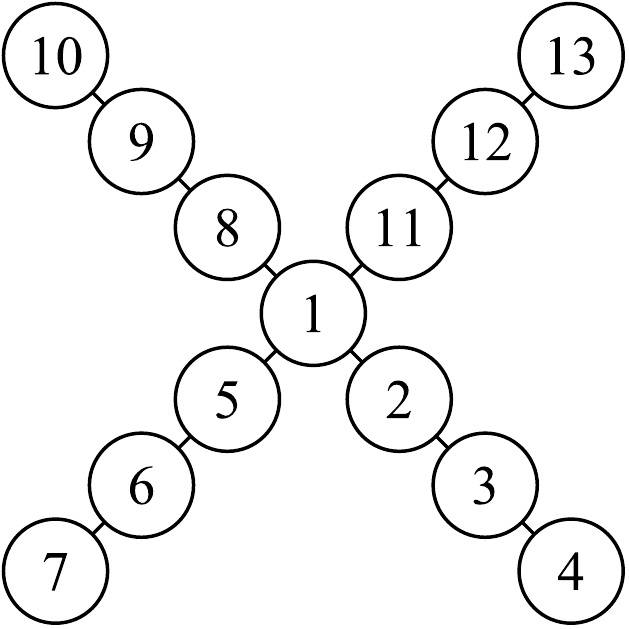}
  \caption{\label{fig:ext_star_large}Starlike networks with (a) $L=2$,
    $R=4$; (b) $L=3$, $R=3$; (c) $L=3$, $R=4$.}
\end{figure}

\begin{figure}
  \centering 
  \widfig{0.32}{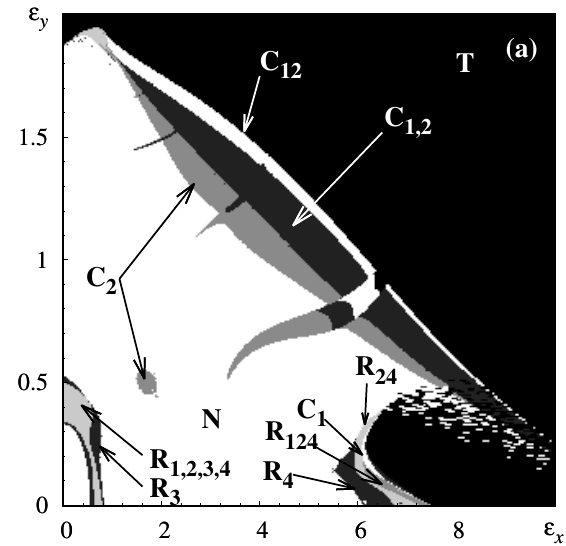}
  \widfig{0.32}{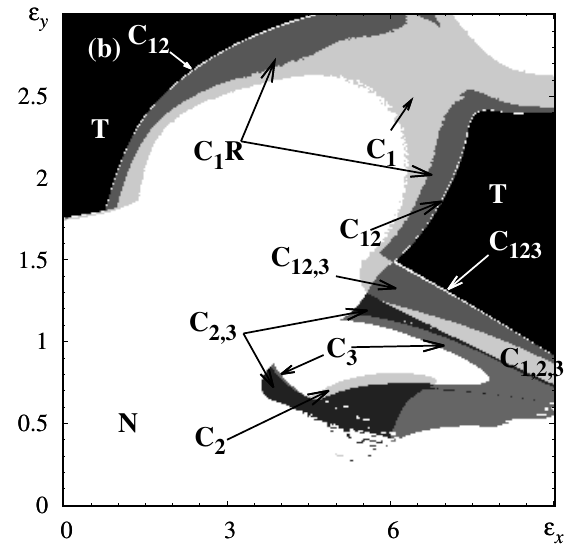}
  \widfig{0.32}{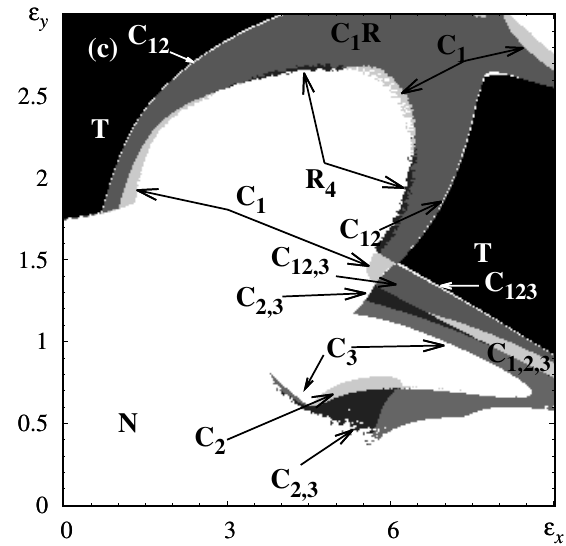}
  \caption{\label{fig:charts_large}Synchronization charts for starlike
    networks with (a) $L=2$, $R=4$; (b) $L=3$, $R=3$; (c) $L=3$,
    $R=4$, see Fig.~\ref{fig:ext_star_large}. Observe that the number
    of rays $R$ does not influence much the overall structure of the
    chart, compare panel (a) with Fig.~\ref{fig:charts}(c) and panel
    (b) with (c). The ray length $L$ however changes the picture
    dramatically.}
\end{figure}

Let us now consider stars whose rays are longer and their number is
higher, see
Fig.~\ref{fig:ext_star_large}. Figure~\ref{fig:ext_star_large}(a)
shows the star with $R=4$, $L=2$, i.e., it has one more ray in
comparison with Fig.~\ref{fig:ext_star}. Similarly to the previous
case natural frequencies of the oscillators, see
Eq.~\eqref{eq:netw_vdp}, strongly decay from the hub to the periphery:
\begin{equation}
  \omega_1=2.5,\;
  \omega_{2,4,6,8}=\{1.475,1.525,1.425,1.575\},\;
  \omega_{3,5,7,9}=\{0.975,1.025,0.925,1.075\}.
\end{equation}
Bifurcation parameters correspond to Fig.~\ref{fig:charts}(c), see
Eq.~\eqref{eq:charts_c}:
\begin{equation}
  \mu_{1}=3,\; \mu_{2,4,6,8}=2,\; \mu_{3,5,7,9}=1.
\end{equation}

Figure~\ref{fig:charts_large}(a) shows the chart of synchronization
for this star. The charts for $R=3$, $L=2$ in Fig.~\ref{fig:charts}(c)
and for $R=4$, $L=2$ in \ref{fig:charts_large}(a) are very similar to
each other. Areas of total synchronization \regmark{T} are almost
identical. In both cases transition to it occurs via circular
synchronization \regmark{C_2}, \regmark{C_{1,2}} and
\regmark{C_{12}}. The belt of radial synchronization
\regmark{R_{1,2,3}} located near the origin in
Fig.~\ref{fig:charts}(c) coincide with the belt \regmark{R_{1,2,3,4}}
in Fig.~\ref{fig:charts_large}(a). Finally, the convexity of
\regmark{T} in bottom right parts in both cases are surrounded by a
border of \regmark{C_1} and by a few \regmark{R}-areas. Thus we
observe that the global structure of synchronization charts is rather
insensitive to the number of the rays.

Increase of the ray length $L$ results in dramatic
changings. Figs.~\ref{fig:ext_star_large}(b) and (c) show two stars
with $L=3$ whose numbers of rays are $R=3$ and $R=4$, respectively.
We set the following parameters for $R=3$:
\begin{gather}
  \label{eq:prm_r3l3_omega}
  \omega_1=4,\;
  \omega_{2,5,8}=\{2.5, 2.45, 2.55\},\;
  \omega_{3,6,9}=\{1.6, 1.55, 1.65\},\;
  \omega_{4,7,10}=\{0.7,0.65,0.75\},\\
  \label{eq:prm_r3l3_mu}
  \mu_1=3,\;
  \mu_{2,5,8}=2,\;
  \mu_{3,6,9}=1,\;
  \mu_{4,7,10}=0.5,
\end{gather}
and for $R=4$:
\begin{gather}
  \begin{gathered}
    \omega_1=4,\; \omega_{2,5,8,11}=\{2.475,2.525,2.425,2.575\},\;\\
    \omega_{3,6,9,12}=\{1.575,1.625,1.525,1.675\},\;
    \omega_{4,7,10,13}=\{0.675,0.725,0.625,0.775\},
  \end{gathered}\\
  \mu_1=3,\;
  \mu_{2,5,8}=2,\;
  \mu_{3,6,9}=1,\;
  \mu_{4,7,10}=0.5.
\end{gather}

Figs.~\ref{fig:charts_large}(b) and (c) represent the synchronization
charts for $L=3$, $R=3$ and $L=3$, $R=4$, respectively. First of all
notice that the these two charts are strongly different from the case
$L=2$, cf. with Figs.~\ref{fig:charts}(c) and
\ref{fig:charts_large}(a). Moreover observe their high similarity:
compare areas \regmark{C_{2,3}}, \regmark{C_2}, \regmark{C_3},
\regmark{C_{1,2,3}}, \regmark{C_{12,3}},\regmark{C_{12}}
\regmark{C_{123}}, and, finally \regmark{T}. Unlike the case $L=2$,
large \regmark{R} areas are absent. Instead there are areas
\regmark{C_{1}R} where synchronized near-hub nodes are also
synchronized with several rays.

As discussed above, at $L=2$ the scenario of transition to the total
synchronization via remote synchronization includes regimes
\regmark{C_{2}}, \regmark{C_{1,2}}, and \regmark{C_{12}}, see
Figs.~\ref{fig:charts}(c) and \ref{fig:charts_large}(a). In other
words, far ends get synchronized first, then near-hub nodes also become
synchronous, then these two clusters merges, and, finally the hub also
get involved so that the whole network oscillates synchronously. This
scenarios survives at $L=3$ with an obvious modification due to longer
rays.  In the right parts of the charts in
Figs.~\ref{fig:charts_large}(b) and (c) we observe that far ends get
synchronized first, area \regmark{C_{3}}. Then goes area
\regmark{C_{2,3}}, i.e., there are two independent clusters including
far ends and intermediate nodes. Subsequent area is
\regmark{C_{1,2,3}} which means that all peripheral nodes belong to
circular clusters and they are not synchronized with each other. Area
\regmark{C_{12,3}} is located further. Here near-hub nodes remain
independent, while far ends and intermediate nodes are merged into a
single cluster. The final area before the total synchronization is a
very thin stripe of \regmark{C_{123}}. It represents the
synchronization of all peripheral nodes.

Altogether, comparison of the charts in Figs.~\ref{fig:charts} and
\ref{fig:charts_large} allows to conclude that the ray length $L$
dramatically influences the synchronization scenario in extended
starlike networks, whereas higher the ray number $R$ merely brings
more fine details in the picture. Moreover, the increase of $L$
results in enrichment of the variety of manifestations of the remote
synchronization. In particular, at $L=3$ sole far ends get
synchronized being connected by non-synchronized five node paths.

\section{\label{sec:cut_star}Asymmetric star}

\begin{figure}
  \centering
  a)~\includegraphics[scale=0.75]{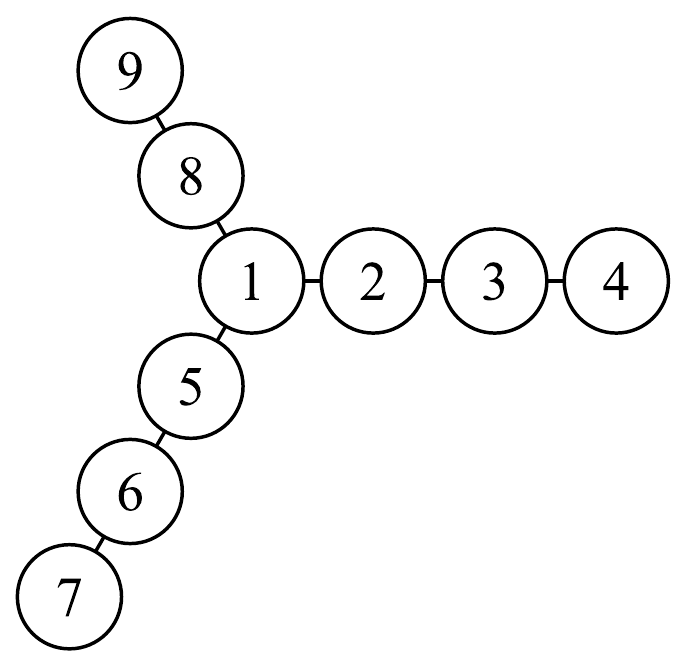}
  \ 
  \widfig{0.32}{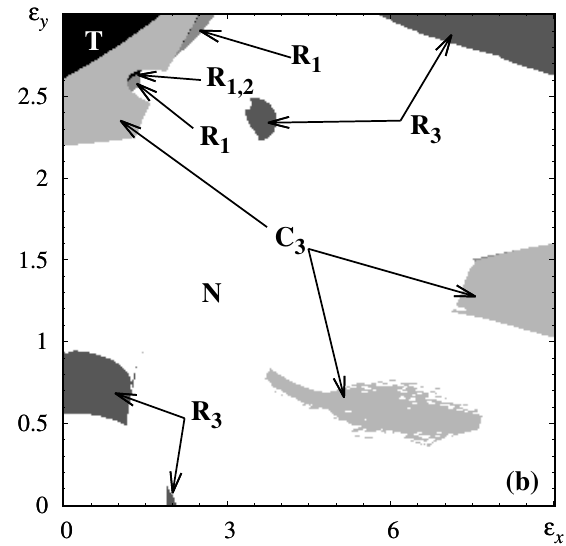}
  \caption{\label{fig:cut_star}Asymmetric star, panel (a), cf. with
    Fig.~\ref{fig:ext_star_large}(b), and corresponding
    synchronization chart, panel (b), cf. with
    Fig.~\ref{fig:charts_large}(b). Cutting off one node results in
    total reorganization of the chart. Circular clusters mainly vanish
    in favor of radial ones. Nevertheless, long-range remote
    synchronization survives, see areas \regmark{C_3}.}
\end{figure}

Let us now consider an asymmetric star. To perform a comparison we
take a network shown in Fig.~\ref{fig:ext_star_large}(b), whose
parameters are given by Eqs.~\eqref{eq:prm_r3l3_omega}
and~\eqref{eq:prm_r3l3_mu}, and cut off its last node 10. The
resulting asymmetric network is shown in Fig.~\ref{fig:cut_star}(a).

Figure~\ref{fig:cut_star}(b) represents a synchronization chart for
the asymmetric star, and the chart for the corresponding uncut star is
in Fig.~\ref{fig:charts_large}(b). We observe that the removing of one
peripheral node results in the total reorganization of the
pictures. Fully symmetric case in Fig.~\ref{fig:charts_large}(b) is
dominated by circular clusters, while radial ones emerge only together
with \regmark{C_1}.

Cutting the node results in the vanish of the most of circular
clusters and emerging of areas of radial clusters instead. It occurs
because nodes involved in the circular clusters are located
symmetrically with respect to rotations of the network around the
hub. The lack of topology symmetry results in the destruction of these
symmetry related clusters. Also this is the case for the total
synchronization regime that shrinks when the topology symmetry is
damaged: compare areas \regmark{T} in Figs.~\ref{fig:charts_large}(b)
and \ref{fig:cut_star}(b).  The radial clusters, on the contrary, are
not related to the topology symmetry so that its lack accompanied with
the vanish of circular clusters benefits to them.

Notice that the long-range remote synchronization does not fully
vanish in the asymmetric case. In Fig.\ref{fig:cut_star}(b) we can see
three large areas of clusters \regmark{C_3}. For the considered
network, see Fig.~\ref{fig:cut_star}(a), it means that nodes 4 and 7
are synchronized while other oscillate independently. For the
symmetric stars we discussed above the scenario of transition to the
total synchronization via one by one formation of circular clusters
from far ends to the hub. In the asymmetric case we can also observe
similar situation. As follows from Fig.~\ref{fig:cut_star}(b) if
$\epsilon_y$ grows at constant small $\epsilon_x$ far ends get
synchronized first and, after that, the whole network become
synchronous. On the boundary between areas \regmark{C_3} and
\regmark{T} there is very narrow stripe of other circular
clusters. However, we do not highlight it in the plot since it is
barely visible.

\section{\label{sec:amp}Amplitude equations}

To verify typical nature of circular and radial cluster
synchronizations, consider amplitude equations for the
network~\eqref{eq:netw_vdp}. They are also known as Stuart-Landau
equations~\cite{PikRosKurtSyn}. Their derivation in our case is
basically straightforward. According to the standard
routine~\cite{PikRosKurtSyn}, it is assumed that solution can be
represented as a fast harmonic oscillation with slow amplitude
modulation. Taking identical for all nodes fast component with
frequency $\omega_1$, we obtain the assumed solution:
\begin{equation}
  \label{eq:fast_and_slow}
  x_n(t)=A_n\mye^{\ramuno\omega_1t}+A^*_n\mye^{-\ramuno\omega_1t},
\end{equation}
where $A_n\equiv A_n(t)$ is a slow varying complex amplitude and
$A^*_n$ is its complex conjugation. After substitution
Eq.~\eqref{eq:fast_and_slow} to Eq.~\eqref{eq:netw_vdp} we apply the
standard auxiliary condition
\begin{equation}
  \dot A_n\mye^{\ramuno\omega_1t}+\dot A^*_n\mye^{-\ramuno\omega_1t}=0,
\end{equation}
and average resulting equations over fast components. Finally, we
rescale time as $t\to2t$ to eliminate insignificant factor 2, and
obtain a sought network of amplitude equations:
\begin{equation}
  \label{eq:netw_amp}
  \dot A_n=
  \left(\mu_n+\ramuno\frac{\omega_n^2-\omega_1^2}{\omega_1}\right)A_n
  -A_n|A_n|^2+\left(\epsilon_y-\ramuno\frac{\epsilon_x}{\omega_1}\right)
  \left(\sum_{j=1}^N\frac{a_{nj}}{k_n}A_j-A_n\right).
\end{equation}
At $\epsilon_x=0$ and $\mu_1=\mu_2=\cdots=\mu_n=\mu$
Eq.~\eqref{eq:netw_amp} turns to the network used by Bergner et
al.~\cite{RemSyn1} to study the remote synchronization.

\begin{figure}
  \centering
  \widfig{0.32}{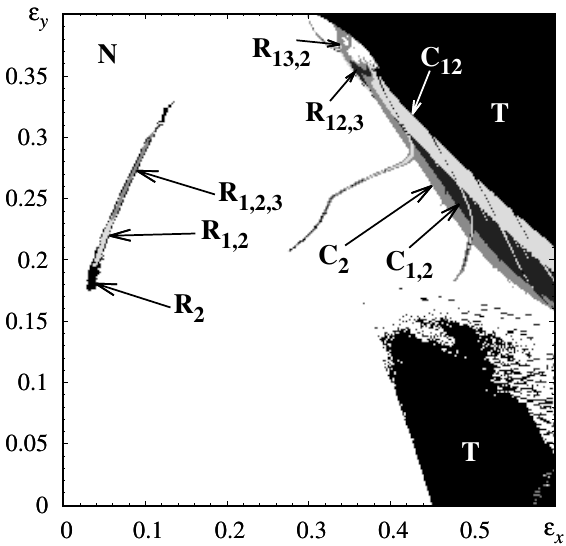}
  \caption{\label{fig:amp_eqn}Synchronization chart for amplitude
    equations~\eqref{eq:netw_amp} on the star at $L=2$, $R=3$, see
    Fig.~\ref{fig:ext_star}. Overall structure of the chart is similar
    to that one for van der Pol oscillators, cf. with
    Fig.~\ref{fig:charts}(c).}
\end{figure}

We consider a starlike network of amplitude
equations~\eqref{eq:netw_amp} with $L=2$, $R=3$, see
Fig.~\ref{fig:ext_star}. Parameters are taken with noticeable
gradients both in $\omega$ and $\mu$:
\begin{gather}
  \label{eq:amp_freq}
  \omega_1=1,\;
  \omega_{2,4,6}=\{0.8 ,\; 0.75,\; 0.85\},\;
  \omega_{3,5,7}=\{0.6 ,\; 0.55,\; 0.65\},\\
  \label{eq:amp_mus}  
  \mu_1=1,\;
  \mu_{2,4,6}=0.8,\;
  \mu_{3,5,7}=0.6.
\end{gather}
Synchronization chart is shown in Fig.~\ref{fig:amp_eqn}. Analogous
parameter set with noticeable gradients, see Eqs.~\eqref{eq:freq_vals}
and \eqref{eq:charts_c}, is employed for the original
system~\eqref{eq:netw_vdp} in Fig.~\ref{fig:charts}(c). We can see
that the charts in Figs.~\ref{fig:amp_eqn} and \ref{fig:charts}(c)
have many similar features. In both cases areas \regmark{T} of total
synchronization are located in the top right parts and separated from
areas \regmark{N} where no synchronization occurs by stripes of
circular clusters. First goes \regmark{C_2}, i.e., only far ends are
synchronized, then near-hub nodes also become synchronous without
synchronization with the far ends, \regmark{C_{1,2}}, after that both
circular clusters become synchronous, \regmark{C_{12}}, and, finally,
total synchronization takes place, \regmark{T}. In both cases these
stripes are furrowed by the rib-like structures. Also there are large
bulges of total synchronization in the bottom right parts of the
charts. Finally, both charts contains islands of radial
synchronization that, however, are not located in somehow similar
ways.

It is well known that the considered amplitude equations are universal
models for a wide class of self-oscillators with supercritical Hopf
bifurcation. Thus, observed similarity of charts in
Fig.~\ref{fig:amp_eqn} and \ref{fig:charts}(c) indicates that circular
clusters and long-range remote synchronization as well as radial
clusters are typical and robust phenomena of starlike networks of
self-oscillators of mentioned type.

\section{Conclusion}

In this paper we considered an extended starlike network of
nonidentical self-oscillators in presence of both reactive and
dissipative coupling. These networks demonstrated two types of
synchronization clusters which were radial and circular.  Radial
synchronization cluster involves all nodes along a ray, while circular
synchronization cluster consists of nodes from all rays located on the
same distance form the hub. By the nature, circular clusters are the
long-range version of so called remote synchronization. Their feature
is that the involved nodes do not have a direct connections and
interact only through intermediary nodes.

Various forms of long-range remote synchronization can be treated as
elements of scenario of transition to the total synchronization of the
network. We observed that the far ends of rays synchronized
first. Then more circular clusters appeared involving closer to hub
nodes. Subsequently the clusters merged and, finally, all network
became synchronous.

Behavior of the extended starlike networks was found to be strongly
determined by the ray length. Its increment resulted in dramatic
rearrangement of the considered synchronization charts. Varying of the
number of rays, however, affected basically the fine details of the
charts while their coarse grain structure remained unaltered.

Furthermore, we observed that the dynamics essentially depends on the
underlying topology symmetry. It can be treated as one more
manifestation of the dependence on the ray length. When one ray
terminating node was removed, so that the star became asymmetric, the
synchronization chart changed very seriously. Most of circular
clusters, related by their nature to the network symmetry, vanished,
while more radial clusters appeared instead. Nevertheless, asymmetry
did not totally killed the long-range remote synchronization.  The
chart for asymmetric star contained large ares where the far end nodes
were synchronized while all others oscillated independently.

We considered networks of van der Pol oscillators and corresponding
amplitude equations that are known as a universal model for a wide
class of self-oscillators. Inspected synchronization charts for this
two types of dynamical systems showed high qualitative similarity:
circular and radial clusters were detected in both cases, and moreover
overall structures of the charts were mainly similar. This is a
serious argument in favor of typical nature of the reported phenomena
for self-oscillators.

To observe the reported variety of phenomena the natural frequencies
of the oscillators were taken decaying from the hub node along
rays. The nodes located on the same distances from the hub had close
but also nonidentical natural frequencies. Multiple simulations
indicated that this is a sufficient condition for the circular
clusters to appear. It agrees with the paper by Bergner et
al.~\cite{RemSyn1} who study a remote synchronization choosing natural
frequencies in the similar way. Tried bifurcation parameters were
totally identical as well as decaying along rays. Circular clusters
appeared in both cases, however the second choice was more preferable
since the areas of their existence on a synchronization chart became
larger. As for the radial clusters, they obviously prefer identical or
close parameters along rays. However we observed that they still could
emerge when the parameters were sufficiently different. These
considerations of the parameters choice are rather empirical. More
rigorous conditions should be the subject of further studies.

This work was partially (P.V.K.) supported by RFBR Grant
No. 16-02-00135

\end{document}